\documentclass[twoside]{ae100prg}
\usepackage{graphicx}
\usepackage[breaklinks]{hyperref}
\usepackage{booktabs}

\usepackage{amsmath,amsthm,wasysym,amssymb,amsfonts,cite,enumerate,rotating,graphicx}
\usepackage{bm} 
\usepackage[dvips]{color}

\newtheorem{prop}{Proposition}

\def \bu {\mbox{\boldmath{$u$}}}
\newcommand{\pa}{\partial}
\def\d{\mathrm{d}}
\def \bg {\mbox{\boldmath{$g$}}}

\begin{document}

\title{Electric and magnetic Weyl tensors in higher dimensions}

\author{S. Hervik$^1$, M. Ortaggio$^2$, L. Wylleman$^{3}$}

\address{$^1$ Faculty of Science and Technology, University of Stavanger, N-4036 Stavanger, Norway}
\address{$^2$ Institute of Mathematics, Academy of Sciences of the Czech Republic, \v Zitn\' a 25, 115 67 Prague 1, Czech Republic}
\address{$^3$ Faculty of Applied Sciences TW16, Ghent University, Galglaan 2, 9000 Gent, Belgium}

\email{sigbjorn.hervik@uis.no, ortaggio@math.cas.cz, lode.wylleman@ugent.be}

\begin{abstract}
Recent results on purely electric (PE) or magnetic (PM) spacetimes in $n$ dimensions are summarized. These include: Weyl types; diagonalizability; conditions under which direct (or warped) products are PE/PM. 

\end{abstract}

\section{Definition and general properties}

The standard decomposition of the Maxwell tensor $F_{ab}$ into its electric and magnetic parts $\vec E$ and $\vec B$ 
with respect to (wrt) an observer (i.e., a unit time-like vector $\bu$) can be extended to any tensor in an $n$-dimensional spacetime \cite{Senovilla00,Senovilla01,HerOrtWyl12}. Here we summarize the results of \cite{HerOrtWyl12} about 
the Weyl tensor, and the connection with the null alignment classification \cite{Milsonetal05,Coleyetal04}. 

Consider the $\bu$-orthogonal projector $h_{ab}=g_{ab}+u_au_b$. The ``electric'' and ``magnetic'' parts of $C_{abcd}$ can be defined, respectively, as 
\cite{HerOrtWyl12}
\begin{eqnarray}
	&&(C_+)^{ab}{}_{cd}=h^{ae}h^{bf}h_c{}^gh_d{}^h C_{efgh}+4u^{[a}u_{[c}C^{b]e}{}_{d]f}u_eu^f ,\label{C+}\\
	&&(C_-)^{ab}{}_{cd}=2h^{ae}h^{bf}C_{efk[c}u_{d]}u^k+2u_ku^{[a}C^{b]kef}h_{ce}h_{df} . \label{C-}
\end{eqnarray}
These extend the well-known 4D definitions \cite{Matte53,Stephanibook}. 
In any orthonormal frame adapted to $\bu$ the electric [magnetic] part
accounts for the Weyl components with an even [odd] number of indices $u$. {\em At a spacetime point (or region) the Weyl tensor is called ``purely electric [magnetic]'' (from now on, PE [PM]) wrt $\bu$ if $C_-=0$ $[C_+=0]$}. The corresponding spacetime is also called PE $[$PM$]$. Several conditions on PE/PM Weyl tensors follow.

\begin{prop}[Bel-Debever-like criteria \cite{HerOrtWyl12}] 
A Weyl tensor $C_{abcd}$ is: \newline (i) PE wrt $\bu$ iff $u_ag^{ab}C_{bc[de}u_{f]}=0$; (ii) PM wrt $\bu$ iff $u_{[a}C_{bc][de}u_{f]}=0$.
\label{prop_Bel-Debever}
\end{prop}

\begin{prop}[Eigenvalues \cite{HerOrtWyl12}] 
\label{th_Weyl_PE} 
A PE [PM] Weyl operator\footnote{In the sense of the Weyl operator approach of \cite{ColHer09} (see also \cite{Coleyetal12}).} is diagonalizable, and possesses only real $[$purely imaginary$]$ eigenvalues. Moreover, a PM Weyl operator has at least $\frac{(n-1)(n-4)}{2}$ zero eigenvalues.
\end{prop}

\begin{prop}[Algebraic type \cite{HerOrtWyl12}] 
\label{PE_PM_types}
A Weyl tensor which is PE/PM wrt a certain $\bu$ can only be of type G, I$_i$, D or O. 
In the type I$_i$ and D cases, 
the second null direction of the timelike plane spanned by $\bu$ and any WAND is
also a WAND (with the same multiplicity). Furthermore, a type D Weyl tensor is PE iff it is type D$($d$)$, and PM
iff it is type D$($abc$)$.
\end{prop}

\begin{prop}[Uniqueness of $\bu$ \cite{HerOrtWyl12}] 
\label{prop uniqueness}
A PE $[$PM$]$ Weyl tensor is PE $[$PM$]$ wrt: (i) a unique $\bu$ $($up to sign$)$ in the type I$_i$ and $G$ cases; (ii) any $\bu$ belonging to the space spanned by all double WANDs $($and only wrt such $\bu$s$)$ in the type D case $($noting also that if there are more than two double WANDs the Weyl tensor is necessarily PE $($type D$($d$))$ {\rm \cite{Wylleman12}}$)$.
\end{prop}

\section{PE spacetimes}

\begin{prop}[$\!\!$\cite{HerOrtWyl12}] 
 \label{prop_shear_twist_free}
All spacetimes admitting a shearfree, twistfree, unit timelike vector field $\bu$ are PE wrt $\bu$. In coordinates such that $\bu=V^{-1}\pa_t$, the line-element reads
\begin{equation}
    \d s^2=-V(t,x)^2\d t^2 + P(t,x)^2\xi_{\alpha\beta}(x)\d x^\alpha \d x^\beta.
    \label{sheartwistfree}
\end{equation}
\end{prop}

The above metrics include, in particular, direct, warped and doubly warped products with a one-dimensional timelike factor, and thus all {\em static} spacetimes (see also \cite{PraPraOrt07}). For a warped spacetime $(M,\bg)$ with $M=M^{(n_1)}\times M^{(n_2)}$, 
one has $\bg=e^{2(f_1+f_2)}\left(\bg^{(n_1)}\oplus \bg^{(n_2)}\right)$, where $\bg^{(n_i)}$ is a metric on the factor space $M^{(n_i)}$ ($i=1,2$) and $f_i$ are functions on $M^{(n_i)}$ ($M^{(n_i)}$ has dimension $n_i$, $n=n_1+n_2$, and $M^{(n_1)}$ is Lorentzian).
\begin{prop}[Warps with $n_1=2$ \cite{PraPraOrt07,HerOrtWyl12}] 
\label{prop: warp 2D factor} 
	A (doubly) warped spacetime with $n_1=2$ is either type O, or type D(d) and PE wrt {any} $\bu$ living in $M^{(n_1)}$; the uplifts of the null directions of the tangent space to $(M^{(n_1)},\bg^{(n_1)})$ are double WANDs of $(M,\bg)$. If $(M^{(n_2)},\bg^{(n_2)})$ is Einstein the type specializes to D(bd), and if it is of constant curvature to D(bcd).
\end{prop}
In particular, all spherically, hyperbolically or plane symmetric spacetimes belong to the latter special case.

\begin{prop}[Warps with $n_1=3$ \cite{PraPraOrt07,HerOrtWyl12}] 
\label{prop_warped3} 
	A (doubly) warped spacetime with $(M^{(n_1)},\bg^{(n_1)})$ Einstein and $n_1=3$ is of type D(d) or O. The uplift of {\em any} null direction of the tangent space to $(M^{(n_1)},\bg^{(n_1)})$ is a double WAND of $(M,\bg)$, which is PE wrt {any} $\bu$ living in $M^{(n_1)}$.
\end{prop}

\begin{prop}[Warps with $n_1>3$ \cite{PraPraOrt07,HerOrtWyl12}] 
\label{prop_warped>3} 
In a (doubly) warped spacetime 
	\begin{enumerate}[(i)]
	\item if $(M^{(n_1)},\bg^{(n_1)})$ is an Einstein spacetime of type D, $(M,\bg)$ can be only of type D (or O) and the uplift of a double WAND of $(M^{(n_1)},\bg^{(n_1)})$ is a double WAND of $(M,\bg)$
	\item if $(M^{(n_1)},\bg^{(n_1)})$ is of constant curvature, $(M,\bg)$ is of type D(d) (or O) and the uplifts of {\em any} null direction of the tangent space to $(M^{(n_1)},\bg^{(n_1)})$ is a double WAND of $(M,\bg)$; $(M,\bg)$ is PE wrt {any} $\bu$ living in $M^{(n_1)}$.
	\end{enumerate}
\end{prop}

\begin{prop}[PE direct products \cite{HerOrtWyl12}] 
 \label{prop_products u-in-M1}
A direct product spacetime $M^{(n)}=M^{(n_1)}\times M^{(n_2)}$ is PE wrt a $\bu$ that lives in $M^{(n_1)}$ iff $\bu$ is an eigenvector of $R^{(n_1)}_{ab}$, and $M^{(n_1)}$ is PE wrt $\bu$.
($\bu$ is then also an eigenvector of the Ricci tensor $R_{ab}$ of $M^{(n)}$, i.e., $R_{ui}=0$.)
\end{prop}

A conformal transformation (e.g., to a (doubly) warped space) will not, of course, affect the above conclusions about the Weyl tensor. There exist also direct products which are PE wrt a vector $\bu$ {\em not} living in $M^{(n_1)}$ \cite{HerOrtWyl12}. 

Also the presence of certain (Weyl) isotropies (e.g., $SO(n-2)$ for $n>4$) implies that the spacetime is PE, see \cite{ColHer09,HerOrtWyl12} for details and examples.

\section{PM spacetimes}

\begin{prop}[PM direct products \cite{HerOrtWyl12}] 
\label{prop_PMproducts} A direct product spacetime $M^{(n)}=M^{(n_1)}\times M^{(n_2)}$ is PM wrt
a $\bu$ that lives in $M^{(n_1)}$ iff all the following conditions hold (where $R_{(n_i)}$ is the Ricci scalar of $M^{(n_i)}$):
\begin{enumerate}[i)]
	\item $M^{(n_1)}$ is PM wrt $\bu$ and has a Ricci tensor of the form $R^{(n_1)}_{ab}=\frac{R_{(n_1)}}{n_1}g^{(n_1)}_{ab}+u_{(a}q_{b)}$ (with $u^aq_a=0$) 
	\item $M^{(n_2)}$ is of constant curvature and {${n_2 (n_2-1)R_{(n_1)}+n_1(n_1-1)R_{(n_2)}=0}$}.
\end{enumerate}
Further, $M^{(n)}$ is PM Einstein iff $M^{(n_1)}$ is PM Ricci-flat and $M^{(n_2)}$ is flat.
\end{prop}

See \cite{HerOrtWyl12} for explicit (non-Einstein) examples. However, in general PM spacetimes are most elusive. For example, 

\begin{prop}[$\!\!$\cite{HerOrtWyl12}]
	PM Einstein spacetimes of type D do not exist.
\end{prop}

In \cite{HerOrtWyl12} also several results for PE/PM Ricci and Riemann tensors have been worked out, along with corresponding examples. In general, we observe that PE/PM tensors provide examples of {\em minimal tensors} \cite{RicSlo90}. Thanks to the {\em alignment theorem} \cite{Hervik11}, the latter are of special interest since they are precisely the {\em tensors characterized by their invariants} \cite{Hervik11} (cf. also \cite{HerOrtWyl12}).  This in turn sheds new light on the classification of the Weyl tensor \cite{Coleyetal04}, providing a further invariant characterization that distinguishes the (minimal) types G/I/D from the (non-minimal) types II/III/N.

\section*{Acknowledgments}

M.O. acknowledges support from research plan {RVO: 67985840} and research grant no P203/10/0749. 

\section*{References}


\begin{thebibliography}{10}

\bibitem{ColHer09}
Coley, A.  and Hervik, S., ``Higher dimensional bivectors and classification of
  the {W}eyl operator'', {\em Class. Quantum Grav.}, {\bf 27}, 015002, (2009).

\bibitem{Coleyetal12}
Coley, A., Hervik, S., Ortaggio, M.  and Wylleman, L., ``Refinements of the
  {W}eyl tensor classification in five dimensions'', {\em Class. Quantum
  Grav.}, {\bf 29}, 155016, (2012).

\bibitem{Coleyetal04}
Coley, A., Milson, R., Pravda, V.  and Pravdov\'a, A., ``Classification of the
  {W}eyl tensor in higher dimensions'', {\em Class. Quantum Grav.}, {\bf 21},
  L35--L41, (2004).

\bibitem{Hervik11}
Hervik, S., ``A spacetime not characterized by its invariants is of aligned
  type {II}'', {\em Class. Quantum Grav.}, {\bf 28}, 215009, (2011).

\bibitem{HerOrtWyl12}
Hervik, S., Ortaggio, M.  and Wylleman, L., ``Minimal tensors and purely
  electric or magnetic spacetimes of arbitrary dimension'', (2012).
  {\small[\href{http://arxiv.org/abs/1203.3563}{{arXiv:1203.3563
  {\small[gr-qc]}}}]}.

\bibitem{Matte53}
Matte, A., ``Sur de nouvelles solutions oscillatoires de \'equations de la
  gravitation'', {\em Canadian J. Math.}, {\bf 5}, 1--16, (1953).

\bibitem{Milsonetal05}
Milson, R., Coley, A., Pravda, V.  and Pravdov\'a, A., ``Alignment and
  algebraically special tensors in {L}orentzian geometry'', {\em Int. J. Geom.
  Meth. Mod. Phys.}, {\bf 2}, 41--61, (2005).

\bibitem{PraPraOrt07}
Pravda, V., Pravdov\'a, A.  and Ortaggio, M., ``Type {D} {E}instein spacetimes
  in higher dimensions'', {\em Class. Quantum Grav.}, {\bf 24}, 4407--4428,
  (2007).

\bibitem{RicSlo90}
Richardson, R.~W.  and Slodowy, P.J., ``Minimum Vectors for real reductive
  algebraic groups'', {\em J. London Math. Soc.}, {\bf 42}, 409--429, (1990).

\bibitem{Senovilla00}
Senovilla, J. M.~M., ``Super-energy tensors'', {\em Class. Quantum Grav.}, {\bf
  17}, 2799--2841, (2000).

\bibitem{Senovilla01}
Senovilla, J. M.~M., ``General electric-magnetic decomposition of fields,
  positivity and {R}ainich-like conditions'', in Pascual-S\'anchez, J.~F.,
  Flor\'{\i}a, L., San~Miguel, A.  and F., Vicente, eds., {\em Reference Frames
  and Gravitomagnetism}, pp. 145--164, (World Sicentific, Singapore, 2001).

\bibitem{Stephanibook}
Stephani, H., Kramer, D., MacCallum, M., Hoenselaers, C.  and Herlt, E., {\em
  Exact Solutions of {E}instein's Field Equations}, (Cambridge University
  Press, Cambridge, 2003), second edition.

\bibitem{Wylleman12}
Wylleman, L., ``On {W}eyl type {II} or more special spacetimes in higher
  dimensions'', to appear.

\end{thebibliography}

\end{document}